\documentstyle[preprint,aps,epsf]{revtex}
\tightenlines
\begin{document}
\title{Comparison of two non-primitive methods for path integral simulations: 
Higher-order corrections vs. an effective propagator approach}

\author{F. R. Krajewski and M. H. M\"user}
\address{Inst. f. Physik, Johannes Gutenberg-Universit\"at,
        55099 Mainz, Germany}
\date{\today}

\maketitle
\begin{abstract}
Two methods are compared that are used in path integral simulations.
Both methods aim to achieve faster convergence to the quantum limit
than the so-called primitive algorithm (PA).
One method, originally proposed by Takahashi and Imada, is based on a
higher-order approximation (HOA) of the quantum mechanical density operator.
The other method is based upon an effective propagator (EPr).
This propagator is constructed such that it  produces
correctly one and two-particle imaginary time correlation functions
in the limit of small densities even for finite Trotter numbers $P$.
We discuss the conceptual differences between both methods and
compare the convergence rate of both approaches.
While the HOA method converges faster than
the EPr approach, EPr gives surprisingly good estimates of thermal 
quantities already for $P = 1$. Despite a significant improvement with respect
to PA, neither HOA nor EPr
overcomes the need to increase $P$ linearly with inverse temperature.
We also derive the proper estimator for radial distribution functions
for HOA based path integral simulations.
\end{abstract}

\section{Introduction}
Path integral Monte Carlo (PIMC)~\cite{barker79} and
path integral molecular dynamics (PIMD)~\cite{tuckerman93}
have proven useful in the atomistic simulation of quantum  effects
occuring in condensed matter at low temperatures.
The broad range of applications includes among other things
superfluid $^4$He~\cite{ceperley95}, 
isotope effects in crystalline rare gas solids~\cite{barrat89,muser95}
as well as phase transitions with strong quantum mechanical effects
in Josephson junctions~\cite{jacobs87,jacobs88} 
and molecular solids~\cite{runge92,muser96}.
Recently, path integral methods have also been applied to calculate
low-temperature material properties of systems that
are computationally less easily tractable
than Lennard-Jones type systems:
solid germanium~\cite{noya97}, 
crystalline polyethylene~\cite{martonak98}, 
diamond~\cite{herrero01},
silica~\cite{muser01}, and
even Wigner crystals~\cite{candido01}, to name a few.

A disadvantage of PIMC and PIMD is the increase of necessary computing time
$t_{\rm CPU}$ with decreasing temperature $T$ at a given required
accuracy. Using the so-called primitive approximation together
with the most efficient sampling algorithms that completely eliminate
critical slowing down (see Ref.~\onlinecite{tuckerman93} for a thorough
discussion), it is not possible to overcome
$t_{\rm CPU} \propto 1/T$.
Therefore different improvements on the primitive algorithms have been
suggested that render path-integral simulations more efficient.
One may subdivide the bulk of such attempts into three categories
(i) methods that are based on higher-order approximants (HOA) of the 
high-temperature density matrix~\cite{takahashi84}, 
(ii) methods that use effective propagators (EPr) that automatically yield
the proper two-particle behavior~\cite{ceperley95,pollock88}, and
(iii) methods that decompose the Hamiltonian into a harmonic and an anharmonic
contribution before applying the Trotter formula~\cite{giachetti86,cuccoli93}. 
The latter category, also refered to as the effective potential (EP) method,
will not be discussed any further in this study,
mainly due to its unfavorable scaling of $t_{\rm CPU}$ with particle
number $N$. Additionally, harmonic approximations are problematic for
many systems of interest, in particular those involving $^3$He and $^4$He.
Yet, another advantage of HOA and  EPr methods over the EP method
is that the pathological behaviour of the attractive Coulomb potential
is overcome automatically~\cite{pollock88,li87,muser97}.

In this study, we want to compare the convergence
of the HOA method, the EPr method, and the primitive algorithm
for a simple model system. 
As long as the interaction potentials are well-behaved, the convergence 
does generally not depend on the specific form of the potential~\cite{suzuki87}.
The test model system should of course be chosen such that it does not 
favor intrinsically
one approach over the other, e.g., we may not chose a two-particle
system, because then the EPr model would be exact per definition.
As we are not interested in the EP approach, we can chose a simple monoatomic
chain with harmonic next-neighbor coupling. This choice of system
enables us to do the bulk of the calculations analytically so that
statistical error bars are eliminated completely.
We also want to investigate how the Trotter number $P$ necessary to
keep the systematic errors below a well-defined percentage scales
with inverse temperature for the different approaches investigated
in this study.

It should be emphasized that the HOA method and the EPr method are
conceptually different. In an EPr path integral simulation,
one tries to generate radial distribution functions that are in the
quantum limit (at least in a low-density approximation). 
Evaluating observables such as the thermal expectation 
value of the potential energy 
$\langle V_{\rm pot} \rangle$ is done by simply evaluating the operator
of the potential energy at the given distance.
In a HOA path integral simulation, generalized estimators have to be defined 
even for those observables that are orthogonal in real space. 
We will comment on this in more depth in the following section.
In particular we will derive an expression for the HOA estimator of the
radial distribution function, which has not been given hitherto,
which might explain the sparse use of the method in the literature.
The new estimator will be used to calculate the argon argon radial 
distribution function of those atoms in a simple three-dimensional
Lennard-Jones crystal.
The different methods employed in this study will be outlined
in Sec.~II. In this section we will also present a  simplified
approach to the EPr approach, which we call reduced effective propagator
(r-EPr) approach.
The results will be presented in Sec.~III, and a short
summary is given in Sec.~IV.

\section{Methods}
\subsection{Primitive Algorithm}
The primitive algorithm for path integrals is based on Feynman's idea
to represent the partition function of a quantum mechanical point particle
$Z(\beta)$
as a partition function of a classical ring polymer~\cite{feynman65,feynman72}.
The potential $V_{\rm rp}(\{r\})$ of the classical ring polymer
has the form:
\begin{equation}
V_{\rm rp}(\{r\}) = \sum_{t=1}^P \left[ 
  {1\over 2}  {m P^2 \over  \beta^2 \hbar^2} \left(r_t - r_{t+1}\right)^2 
  + V(r_t) \right],
\label{eq:primitive}
\end{equation}
with $\beta = 1/k_B T$. $r_t$ represents the position of monomer $t$ in the 
ring polymer ($r_t = r_{t+P}$), and $V$ is the real (physical) potential. 
$t$ is sometimes interpreted as an imaginary time and
$P$ is commonly called the Trotter number.
The PIMC or PIMD program are then assumed to generate distributions such
that the probability of configuration $\{r\}$ to occur is proportional to
$\exp[-\beta V_{\rm rp}(\{r\})/P]$.
All thermal expectation values of observables diagonal in real space
can be determined directly from the configurations, e.g.,
\begin{equation}
\langle V \rangle = \lim_{P \to \infty} \lim_{M \to \infty}
 {1 \over M P} \sum_{i=1}^M  \sum_{t=1}^P V(r_{t,i}),
\end{equation}
where $r_{t,i}$ is the position of the $t$'th monomer in the
$i$'th Monte Carlo step and $M$ is the number of observations in the Monte
Carlo simulation.
We refer to
Refs.~\onlinecite{barker79,tuckerman93,ceperley95} for further details on the
primitive algorithm.

\subsection{Higher-Order Approximant Method}
\label{sec:HOA}

The HOA method is based on a fourth-order Hermitian Trotter decomposition
of the high-temperature density matrix~\cite{raedt83}.
The decomposition was first applied to continuous degrees of freedom
by Takahashi and Imada~\cite{takahashi84}  as well as by
Li and Broughton~\cite{li87}. The basic idea of the decomposition
is to approximate the high-temperature density matrix 
$\hat{\rho} = \exp(-\beta \hat{H}/P)$ with
\begin{equation}
\hat{\rho} \approx e^{-\beta \hat{V}/2P} e^{-\beta \hat{T}/2P}
             e^{-\beta \hat{V}_{\rm cor} / P}
             e^{-\beta \hat{T}/2P} e^{-\beta \hat{V}/2P}
\label{eq:decomp}
\end{equation}
where $\hat{H} = \hat{T} + \hat{V}$ corresponds to the Hamiltonian and
$V_{\rm cor} = \beta^2 [[\hat{V},\hat{T}],\hat{V}]/ 24 P^2$ is a correction
term.
$\hat{T}$ and $\hat{V}$ are usually chosen to be kinetic and 
potential energy, respectively. For this decomposition, the 
correction energy $V_{\rm cor}$ can be written as
\begin{equation}
V_{\rm cor} = {\beta^2 \hbar^2 \over 24 P^2 }
\sum_{n=1}^N {1\over m_n} \left( \nabla_n V \right)^2,
\label{eq:correction_v}
\end{equation}
where $m_n$ corresponds to the mass of particle $n$.
Owing to the temperature-dependence of $V_{\rm cor}$, thermal
expectation values of observables $O$ have to be reevaluated with respect
to the primitive algorithm, e.g., averages of functions diagonal in real 
space read
\begin{eqnarray}
\langle \hat{O} \rangle & = & 
{1\over MP} \lim_{P \to \infty} \lim_{M \to \infty} 
 \left[ \sum_{i=1}^M \sum_{t=1}^P  \Bigg\{ O(\{r\}) \right.  \nonumber \\ 
& & \left.\left.
+ \sum_{n=1}^N {\beta^2 \hbar^2 \over 12 m_n P^2} (\nabla_n V) 
(\nabla_n O) \right\} \right],
\label{eq:cor_term}
\end{eqnarray}
where $V$ on the right-hand side of Eq.~\ref{eq:cor_term} only 
includes the original potential and not the correction term $V_{\rm cor}$.
The accuracy of the HOA method outlined above allows one to determine
thermal expectation values with a leading correction of $1/ P^4$,
while the primitive algorithm has leading corrections in the
order of $1/ P^2$\,\,\,~\cite{takahashi84,li87,suzuki87}.

Eq.~(\ref{eq:cor_term}) allows one easily to find the estimator for the
potential energy to be $V + 2V_{\rm cor}$, see 
Refs.~\onlinecite{takahashi84,li87}
for further details on the calculation of thermal expectation values. 
To our knowledge, however, it is has not yet been discussed that even the 
estimator for radial distribution functions $g(r)$ needs to be altered
with respect to the primitive approach, for which  the estimator can be written
as
\begin{equation}
g_{\rm prim}^{\rm estim}(r) 
\propto 
\delta(r-\mid {\bf r}_{t,i,n} - {\bf r}_{t,i,n'} \mid ) \, 
\, / \, r^2 \, ,
\label{eq:prim_rad}
\end{equation}
${\bf r}_{t,i,n}$ denoting the  position of particle $n$ of the 
$t$'s monomer in ring polymer (particle)  $n$.
Applying Eq.~(\ref{eq:cor_term}) to Eq.~(\ref{eq:prim_rad}) leads to a
shift of the estimator for the distance   between particle $n$ and $n'$.
Simply applying Eq.~(\ref{eq:cor_term}) to the operator for the
square of the distance between particle $n$ and $n'$ yields
the estimator $r_{n,n'}^{\rm estim}$ 
for the distance between particle $n$ and $n'$, which
is found to be 
\begin{eqnarray}
r_{n,n'}^{\rm estim}  & = & \left\{
\left(  {\bf r}_{t,i,n} - {\bf r}_{t,i,n'} \right)^2  \right.
\nonumber\\ 
& + & \left.
 {\beta^2 \hbar^2 \over 6 P^2} \left(  {\nabla_n V \over m_n}
-  {\nabla_{n^\prime} V \over m_{n^\prime}} \right) 
\left(  {\bf r}_{t,i,n} - {\bf r}_{t,i,n'} \right) 
\right\}^{1/2} .
\label{eq:shift}
\end{eqnarray}
Thus $r_{n,n'}^{\rm estim}$ should replace $r$ in the argument of the
$\delta$-function on the right hand side of
Eq.~(\ref{eq:prim_rad}) in order to calculate $g(r)$. 
Based on this relation, one may say that the HOA method is not an importance
sampling algorithm in the sense that the probability for a
configuration to occur in the simulation is  proportional to 
the diagonal elements of $\hat{\rho}$ in a real space representation.

\subsection{Effective Propagator Approach}
An alternative non-primitive method to the HOA method is to construct effective
potentials such that the two-particle propagators are reflected accurately,
e.g., it is correct in all orders of $\hbar$.
This effective potential is then used in a multi-particle simulation
and hence produces the proper thermal behaviour in the low-density limit.
No effective estimators have to be defined for observables diagonal in
real space and in this sense, the approach is an importance sampling algorithm.
The effective propagator (EPr) method is particularly useful for
ill-behaved potentials such as the attractive Coulomb
potential~\cite{pollock88}. One may write the high-temperature
two-particle density operator in the following way:
\begin{eqnarray}
& &\langle {\bf r}_1 {\bf r}_2 \mid e^{-\beta \hat{H}/P}
\mid {{\bf r}'}_1 {{\bf r}'}_2 \rangle \,\, \propto \,\,
\exp{\left[-{\beta \over P} V_{\rm eff}\right] }
\nonumber\\  &  & \times
\exp{\left[-{\beta\over P}{mP^2\over 2 \beta^2\hbar^2}
 \left\{ ({\bf r}_{1}-{\bf r}'_{1})^2 + 
         ({\bf r}_{2}-{\bf r}'_{2})^2 \right\}\right]} \, .
\label{eq:eff_prop_def}
\end{eqnarray}
$V_{\rm eff}$ is a function that depends on 
${\bf r}_1, {\bf r}'_1,  {\bf r}_2$, and ${\bf r}'_2$. Therefore the 
interaction can be said to be non-local in imaginary time, e.g., in the
primitive decomposition ${\bf r}_1$ does not couple directly to ${\bf r}'_2$.
The calculation of both the diagonal and the non-diagonal elements in more
than one dimension is not trivial for non-harmonic potentials
and we refer to Ref.~\onlinecite{ceperley95}
for an in-depth discussion of that problem. 
For our one-dimensional model system, however, 
the approach can be simplified significantly, i.e., it can be
solved analytically up to a summation over finite number of terms.
This will be done in the Section~\ref{sec:results}.

\subsubsection{The r-EPr method}
\label{sec:r-EPr}

In general the implementation of the full two-particle pair propagator with 
correct diagonal and non-diagonal elements is difficult because, one 
has to use the two-particle high-temperature density matrix (HTDM)
$ \langle {\bf r}_1 {\bf r}_2 \mid e^{-\beta \hat{H}/P}
\mid {{\bf r}'}_1 {{\bf r}'}_2 \rangle$.
In principle one can evaluate this expression prior to the simulation.
However, this is a difficult task due to the dimensionality of the
matrix, in particular for anharmonic $d = 3$ dimensional systems.
Even if one expresses the HTDM in the center of mass system of the coordinates 
${\bf r}_1, {\bf r}_2, {{\bf r}'}_1, {{\bf r}'}_2$ and even if one uses an 
appropriate orientation of the axes within a molecular-fixed frame,
one is a left with a three dimensional table and transformations between
molecular fixed frames and laboratory system. It is also difficult to find
good fit functions for the effective potential $V_{\rm eff}$.

The main idea of a reduced effective potential approach (r-EPr) is to only 
incorporate those corrections to the primitive decomposition that are local,
e.g.,  those corrections that involve coordinates at different Trotter
indices are neglected. In other words, if one carries out a simulation
at inverse temperature $\beta$ with Trotter number $P$, one uses an
effectively classical potential that reproduces the correct two-particle
distribution function in the low-density limit at inverse temperature 
$\beta/P$.  This approach is similar to an idea suggested by 
Thirumalai et al.~\cite{thirumalai84}, who constructed effective
interaction potentials from the diagonal elements of the 
high-temperature density matrix, see also a related paper by 
Pollock and Ceperley~\cite{pollock84}.

\section{Results}
\label{sec:results}
\subsection{Linear, harmonic chain}
In order to analyze the convergence of primitive algorithm, HOA and EPr method
respectively, we chose a one-dimensional linear chain with harmonic next 
neighbor coupling
\begin{equation}
V = {1\over 2} \sum_{n=1}^N k\, (r_n - r_{n+1})^2.
\label{eq:harmonic}
\end{equation}
Periodic boundary conditions, $r_{N+1} = r_1$ are applied and the masses
$m$ are identical for all atoms.

\begin{figure}[hbtp]
\begin{center} \leavevmode
\hbox{ \epsfxsize=100mm \epsfbox{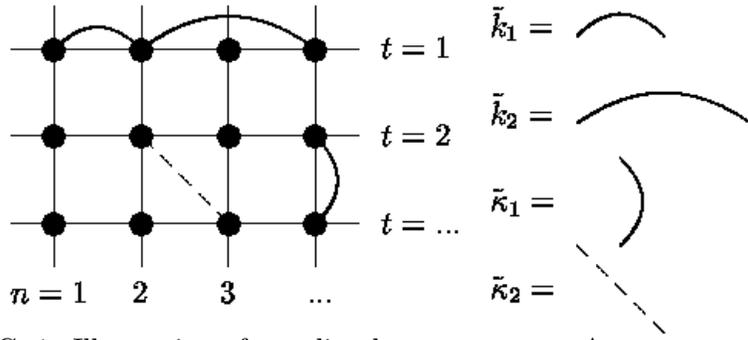} }
 \begin{minipage}{12cm}
  \caption{Illustration of coupling between atoms. Atoms can only oscillate in
   horizontal direction. Vertical springs also act in horizontal direction.
   The straight horizontal lines represent springs between nearest neighbors 
   of stiffness $k$,
   the solid vertical lines springs of stiffness $mk_b^2T^2P^2/\hbar^2$.
  \label{fig:illu} }
 \end{minipage}
 \end{center}
\end{figure}

HOA and EPr invoke correction terms in the potential energy
of the ring polymers with respect to the original expression of
the primitive approach that is given 
in Eq.~(\ref{eq:primitive}) for a one particle problem.
All three approaches can be represented as the limiting case of
a (1+1)-dimensional solid with harmonic coupling between nearest neighbors,
next nearest and next next nearest neighbors. A graphical illustration is
given in Figure~\ref{fig:illu}.

The new effective energy $\tilde{V}_{\rm rp}$ that enters the
Boltzmann factor reads:
\begin{eqnarray}        
\tilde{V}_{\rm rp}   &=&  {1\over2} \sum_{n=1}^N  \sum_{t=1}^P \sum_\pm \left[ \,\,
\left(\kappa+\tilde{\kappa}_1\right) \left(r_{t,n} - r_{t+1,n}\right)^2 \right. \nonumber \\
 && + (k+\tilde{k}_1) (r_{t,n} - r_{t,n+1})^2  +
\tilde{\kappa}_2 (r_{t,n} - r_{t\pm 1,n+1})^2 \nonumber \\
&& \left. + 
\tilde{k}_2 (r_{t,n} - r_{t,n+2})^2 \,\, \right].
\end{eqnarray}
The expression $\tilde{V}_{\rm rp}$ can be diagonalized in terms of
Fourier modes
\begin{equation}
\tilde{r}_{q,\omega}  =  \sqrt{{1\over PN}} \sum_{t=1}^P 
    \sum_{n=1}^N\, r_{t,n} \, e^{2\pi i q n / N} \,
e^{2\pi i \omega t / P}.
\label{eq:fourier}
\end{equation}
It is convenient to introduce $k' = k + \tilde{k}_1$ and
$\kappa' = \kappa + \tilde{\kappa}_1$ with 
$ \kappa = m P^2 /( \beta^2 \hbar^2 ) $ 
in order to reexpress $\tilde{V}_{\rm rp}$
as
\begin{equation}
\tilde{V}_{\rm rp} = {1\over 2} \sum_{\omega=1}^P  \sum_{q=1}^N
k_{q,\omega} \, \mid \tilde{r}_{q,\omega} \mid^2
\end{equation}
with
\begin{eqnarray}
\tilde{k}_{q,\omega} &=&
\kappa^\prime 4 \sin^2 \left( {\pi \over P} \omega \right)
 + k^\prime 4 \sin^2 \left({ \pi \over N} q \right) \nonumber \\
 &&
    + \tilde{\kappa}_2 4 
     \left[ \sin^2 \left( \pi(q/N + \omega/P \right) 
    +       \sin^2 \left( \pi(q/N - \omega/P \right) \right]
    \nonumber \\
 &&
     + \tilde{k}_2 4 \sin^2 \left({\pi \over N} 2q \right)   \,.
\label{eq:all_k}
\end{eqnarray}
The partition function $Z_c$ 
for the classical system illustrated in 
Fig.~\ref{fig:illu} can then  be reduced to
$NP$ Gaussian integrals. $Z_c$ is
proportional to 
\begin{equation}
Z_c \propto \prod_{q=1}^N \prod_{\omega=1}^P
   \sqrt{ \tilde{k}_{q,\omega}}^{\,\,\,-1} \;.
\end{equation}
For the different algorithms we find different 
functions for $\tilde{k}_1, \tilde{k}_2,\tilde{\kappa}_1$,
and $\tilde{\kappa}_2$.
The expressions for these effective coupling coefficients are summarized in 
Table~\ref{tab:results}. The expressions $A$ and $C$ used in 
Table~\ref{tab:results} for the EPr treatment will be given and derived below.

We will now calculate the thermal expectation value of the
potential energy $\langle V_P \rangle$ for a given Trotter number $P$.
One can expect that the convergence rate does not depend on the
observable,  which is why it is sufficient to only investigate
$\langle V_P \rangle$.
The calculations for the three approaches will be separated into three
subchapters. 

\subsubsection{Solution for the primitive Method}

In the primitive method, the usual thermodynamic relationships can be
used in a straightforward way without modification, albeit their 
use can be impractical for technical reasons.  
In order to calculate the thermal expectation
value of $\langle V_P \rangle$, one can use the relation:
\begin{equation}
\label{eq:pot_energy}
\langle V_P \rangle = - {k \over \beta }{d\over d k} \ln (Z_c),
\end{equation}
where all effective coupling coefficients vanish.
This relationship simply follows from the formal expression for the 
quantum mechanical partition function of a linear, monoatomic harmonic chain.
The final expression for $\langle V_P \rangle$ then is:
\begin{equation}
\label{eq:pot_energy_pa} 
\langle V_P \rangle = {k\over 2 \beta} \sum_{q=1}^N \sum_{\omega=1}^P 
{4 \sin^2\left(\pi q/ N\right) \over \tilde{k}_{q,\omega}}\,.
\end{equation}
This equation can be interpreted as follows: Equipartition requires
that $\tilde{k}_{q,\omega} \mid x_{q,\omega} \mid^2$ = $k_B T$.
Hence $\tilde{k}_{q,\omega}$ appears in the denominator. The amount of
{\it real} potential energy in these modes, however, is only 
$\tilde{k}_{q}  \mid x_{q,\omega} \mid^2 / 2$, where $\tilde{k}_{q}$
is the stiffnes that can be associated with the mode $q$ in a 
classical linear chain.

\subsubsection{Solution for the HOA method}

Eq.~(\ref{eq:pot_energy}) also holds for the HOA method, because
the only modification  with respect to PA is that a better
approximant for the high-temperature density matrix is emploied.
However, the correct coupling coefficients $\tilde{k}_1$ and $\tilde{k}_2$
have to be used. The expressions for $\tilde{k}_1$ and $\tilde{k}_2$
stated in Table~\ref{tab:results} are obtained in a straightforward way by 
inserting Eq.~(\ref{eq:harmonic}) into Eq.~(\ref{eq:correction_v}).
This leads to the expression
\begin{eqnarray}
\label{}
\langle V_P \rangle &=&  {k \over 2 \beta } \sum_{q=1}^N \sum_{\omega=1}^P
                     \Bigg\{
                     \left( 1 + {2\over 3 } {\hbar^2 \over m } \left( {\beta \over P } \right)^2 k \right)
                        4 \sin^2 \left( {\pi \over N} q \right) \nonumber \\ 
            &&          -    {2\over 3 } {\hbar^2 \over m } \left( {\beta \over P } \right)^2 k
                          \sin^2 \left( {\pi \over N} 2 q \right) \Bigg\}               
                         \Bigg /   \tilde{k}_{q,\omega}^{\rm HOA}  ,
\end{eqnarray}
where $\tilde{k}_{q,\omega}^{\rm HOA}$ refers to that expression for
$\tilde{k}_{q,\omega}$ in Eq.~(\ref{eq:all_k}) which is obtained by
inserting the HOA values for $\tilde{k}_1$ and $\tilde{k}_2$.
The same result for $\langle V_P \rangle$
could have been obtained by calculating the second moments of
the eigenmodes $\langle \mid \tilde{x}_{q,\omega} \mid^2 \rangle$ from
equipartition. The resulting $\langle \mid \tilde{x}_{q,\omega} \mid^2 \rangle$
could then have been used to calculate the proper HOA potential energy 
estimator $V + 2V_{\rm cor}$. 

\subsubsection{Solution for the EPr method}

In the EPr approach, $Z_c$ does not follow from a Hermitian decomposition
of the high-temperature density matrix. Therefore Eq.~(\ref{eq:pot_energy}) 
can not be emploied for the calculation of $\langle V_P \rangle$.
However, we can use the fact that EPr is an importance sampling method.
Since we can obtain $\langle \mid \tilde{x}_{q,\omega} \mid^2 \rangle$ from
equipartition and since the real potential energy of mode $q$ in a (classical)
linear chain is  proportional to $k\sin^2(\pi q/N)$, it is possible to
use Eq.~(\ref{eq:pot_energy_pa}) where the denominator $\tilde{k}_{q,\omega}$
is taken from the EPr column in Table~\ref{tab:results}. Hence,
\begin{equation}
\label{eq:pot_energy_epr} 
\langle V_P \rangle = {k\over 2 \beta} \sum_{q=1}^N \sum_{\omega=1}^P 
{4 \sin^2\left(\pi q/ N\right) \over \tilde{k}_{q,\omega}^{\rm EPr}}\,.
\end{equation}

%

We are now concerned with the derivation of the expressions for $A$ and $C$ in 
Table~\ref{tab:results}. To do this, one has to consider a dimer in 
which the two atoms are coupled by a harmonic spring of stiffness $k$.
One then transforms the dimer Hamiltonian into the center-of-mass system
with $R = (r_1 + r_2)/2$ the center-of-mass coordinate and 
$\Delta \! R = (r_1 - r_2)$ the reduced distance between the two atoms.
These two coordinates can be associated with two modes, the center of mass mode
with mass $M = 2 m$ and the internal oscillator mode with stiffnes $k$ and 
reduced mass $\mu = m/2$. Both the free particle density matrix and
the high-temperatur density matrix (HTDM)of an oscillator with spring constant
$k$ and mass $\mu$ are known exactly. The free particle HTDM is simply
proportional to $\exp[-MP(R-R')^2/2\beta\hbar^2]$ 
(see Eq.~(\ref{eq:primitive})) while the oscillator's HTDM is given 
by~\cite{feynman72}:
\begin{eqnarray}
\label{HTDM}
 && \rho(\Delta \!R, \Delta \!R',\beta/P) =
\sqrt{{\sqrt{\mu k} \over 2 \pi \hbar \sinh(2 f) }} \, \times
\label{eq:kernel}
      \\
 & &  \exp \Bigg\{ {- \sqrt{\mu k} \over 2 \hbar \sinh(f) }
      \Bigg[ ( \Delta \!R^2 +  \Delta \!R'^2 )  \cosh(f)
    - 2  \Delta \!R  \,  \Delta \! R'\Bigg] \Bigg\} \nonumber \; .
\end{eqnarray}
The prefactor on the right-hand side of Eq.~(\ref{eq:kernel}) provides
an irrelevant offset in $V_{\rm eff}$, which will be neglected in the 
following treatment.
One then needs to transform  the product of the free HTDM and the internal 
oscillator HTDM back into the laboratory system and express the effective 
potential according to Eq.~(\ref{eq:eff_prop_def}).
With the definition of
\begin{equation}
f = {\beta  \over P} \, \hbar \sqrt{ 2 k \over m } 
\end{equation}
we obtain the parameters $A$ and $C$
\begin{eqnarray}
A &=& {2\over f} \,  { \tanh \left( {f \over 2} \right) } \; , 
\label{eq:a_def} \\
C &=& {2\over f^2} \left\{ 1 - {f \over \sinh(f)}  \right\} \, 
\end{eqnarray}
that were introduced in Table~\ref{tab:results}.

\subsubsection{Solution for the r-EPr approach}

According to Sec.~\ref{sec:r-EPr}, we need to know the quantum-mechanical
radial distribution function $g(r)$ at inverse temperature $\beta/P$ 
for the reduced harmonic oscillator in order to construct the r-EPr effective
potential $V_{\rm eff}$.  For this purpose, it is  sufficient to know
the internal energy of the reduced harmonic oscillator,
since $g(r)$ is a simple Gaussian at all temperatures. Hence we need
to find the effective coupling constant $k'$ which generates a second moment
$\langle x^2 \rangle$ in a classical treatment at temperature $PT$
such that the 
the real potential energy $k \langle x^2 \rangle / 2$ corresponds to
the  quantum limit at temperature $PT$. This condition, which defines
$k'$, can be written as
\begin{equation}
\left\langle V\left({\beta/ P}\right) 
\right\rangle_{\rm exact} = {1\over 2}\, {k\over k'} k_B T P,
\end{equation}
where $\langle V(\beta/P) \rangle_{\rm exact}$ is the correct thermal
potential energy for an oscillator at temperature $PT$.
For a harmonmic oscillator, 
$\langle V(\beta/P) \rangle_{\rm exact}$ is half the internal energy $U$,
given by
$U = 0.5 \, \hbar \omega \, \coth(\beta\hbar\omega/2)$
with frequency $\omega = \sqrt{k/\mu} = \sqrt{2k/m}$. We solve for $k'$ and
find $k' = A\,k$.
The parameter $A$ is given in Eq.~(\ref{eq:a_def}) and $\tilde{k}_1 = k(A-1)$
as stated in Table~\ref{tab:results} follows.


\subsubsection{Comparison of the methods}

\begin{figure}[hbtp]
\begin{center} \leavevmode
\hbox{ \epsfxsize=100mm \epsfbox{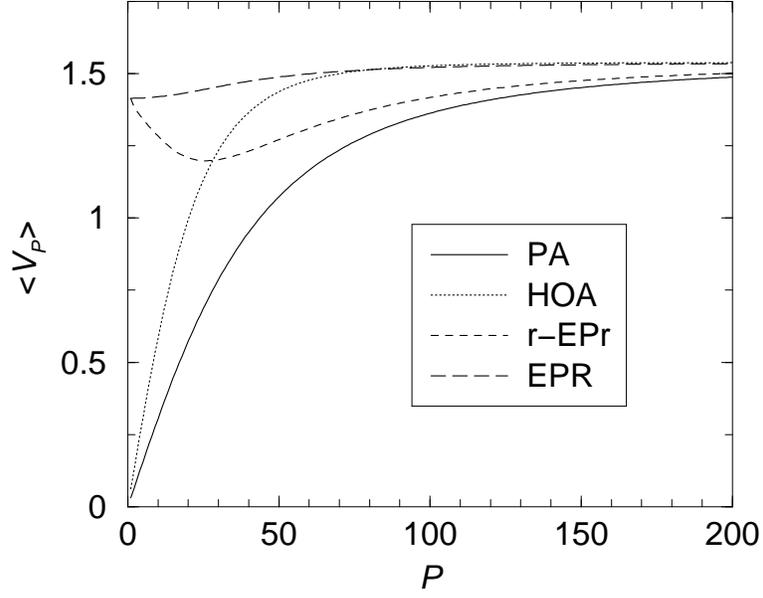} }
 \begin{minipage}{12cm}
   \caption{Thermal expectation value of  the potential energy 
   $\langle  V_P \rangle$ as a function of Trotter number $P$;
   $\beta \hbar \protect\sqrt{k/m} = 64$.}
   \label{conv}
 \end{minipage}
\end{center}
\end{figure}

The main issue of this study is the analysis of the convergence of thermal 
expectation values such as the potential energy $\langle V_P \rangle$ to the
proper quantum limit as a function of Trotter number $P$.
We  consider a linear chain consisting of $N = 5$ atoms and
periodic boundary conditions.
The convergence does not depend on $N$ in a qualitative way.
It is examined at a fixed thermal energy
well below the Debye frequency of the chain, 
namely at inverse temperature $\beta = 64 / (\hbar \sqrt{k/m}) $. 
A linear plot of
$\langle V_P \rangle$ is shown in Fig.~\ref{conv}. 

\begin{figure}[hbtp]
\begin{center} \leavevmode
\hbox{ \epsfxsize=100mm \epsfbox{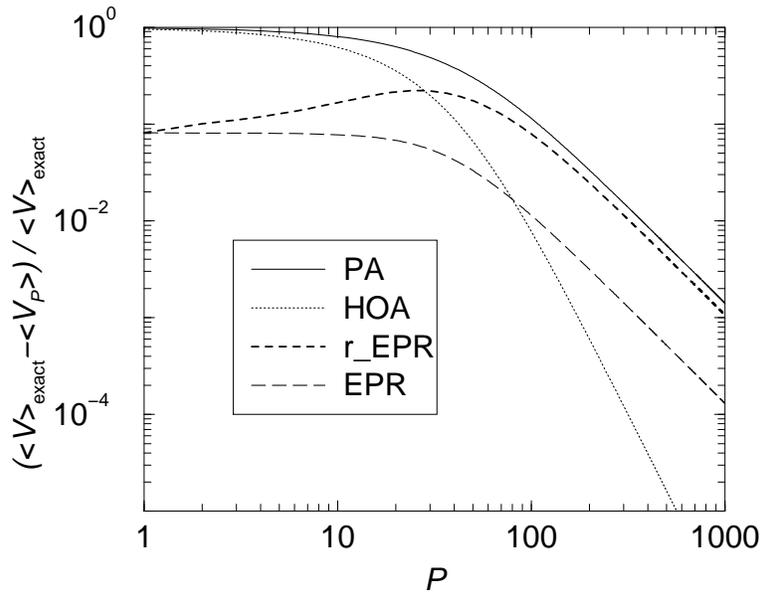} }
 \begin{minipage}{12cm}
  \caption{Relative error of the potential energy for an $N = 5$ 
  chain at $\beta \hbar \omega = 64$ as a function of Trotter number $P$. 
  \label{log_conv} }
 \end{minipage}
 \end{center}
\end{figure}

It can be seen that at $P = 1$ the EPr and the r-EPr method start off with 
estimates that are very close to the quantum limit while PA and HOA start off
with an estimate near the classical value. Upon increasing $P$
the EPr approaches the proper value from below, while for the r-EPr method 
the deviation between estimate and proper result first increases before it
decreases again.
At a Trotter number $P \approx 64$, the HOA method  becomes similarly good as
the EPr approach. In order to address the convergence in a more quantitative 
way, it is convenient to analyze the relative deviation of 
$\langle V_P \rangle$ from the exact value $\langle V_E \rangle$ as a 
function of $P$ in a double logarithmic plot, which is done in 
Fig.~\ref{log_conv}. 

\begin{figure}[hbtp]
\begin{center} \leavevmode
\hbox{ \epsfxsize=100mm \epsfbox{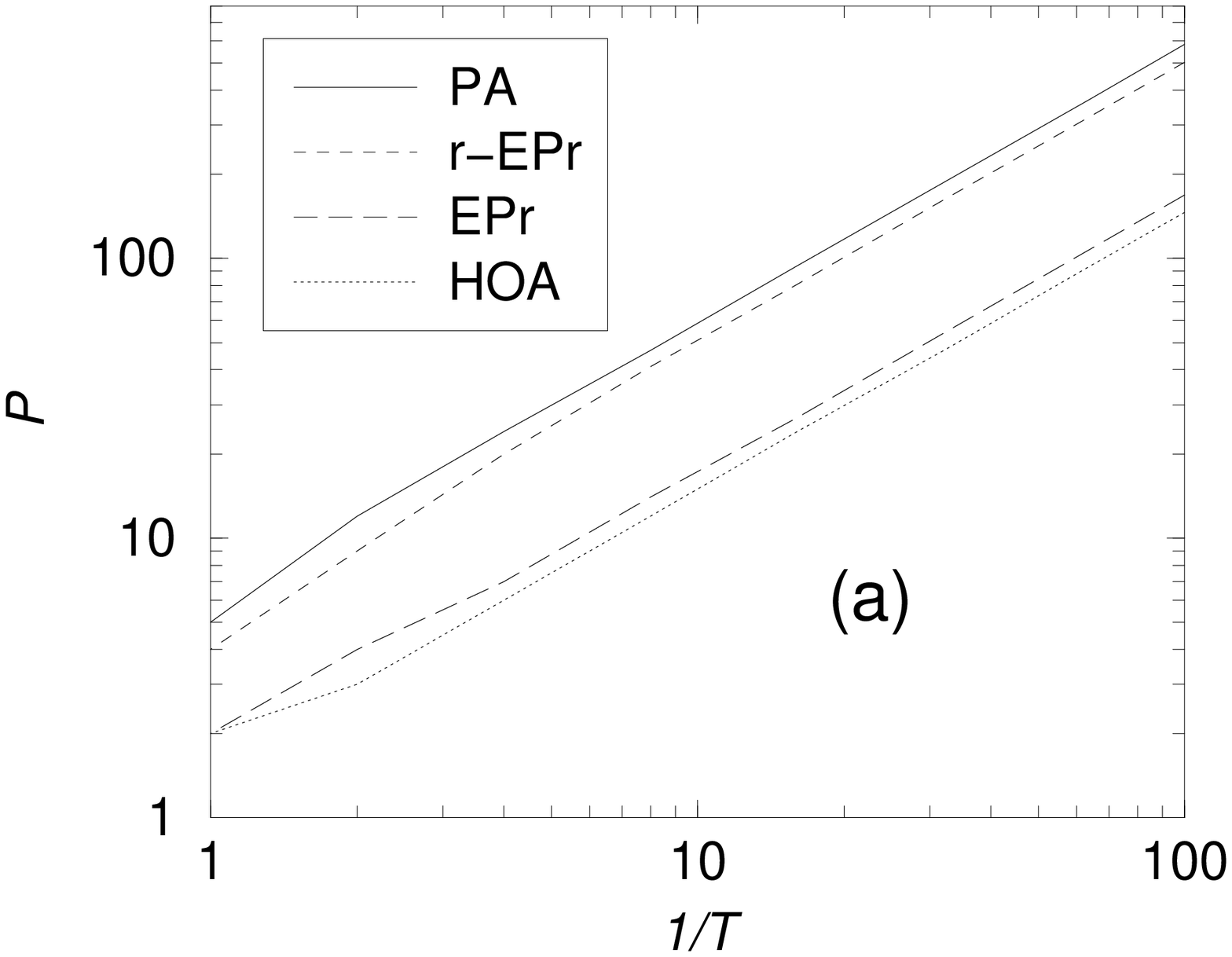} }
\hbox{ \epsfxsize=100mm \epsfbox{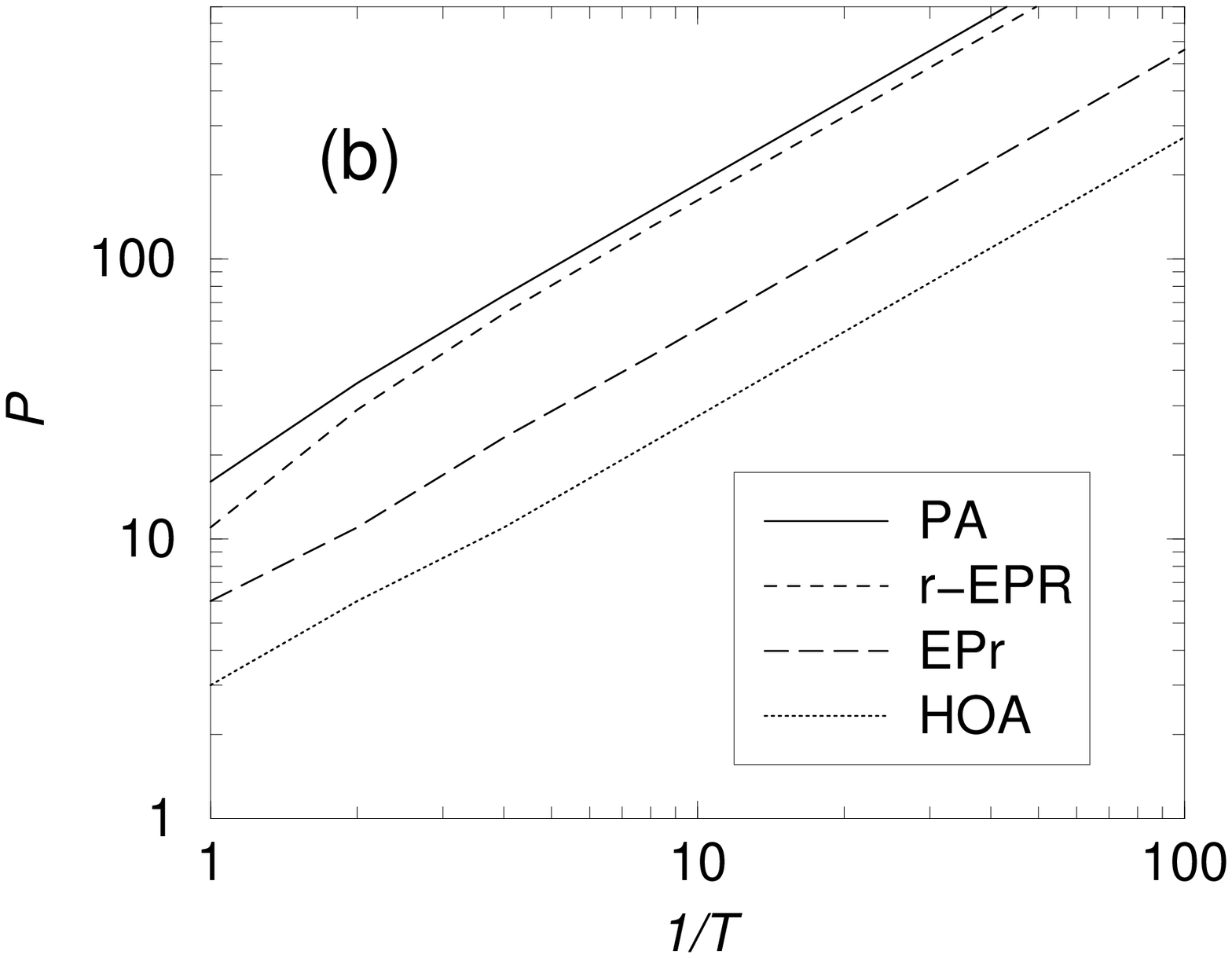} }
 \begin{minipage}{12cm}
  \caption{Necessary Trotter number $P$ to reach a relative
  accuracy of
  a) $10^{-2}$  and b) $10^{-3}$ in the potential energy $\langle V_P \rangle$
   at different inverse temperatures $1/T$ for the linear chain
   consisting of $N = 5$ atoms.
  \label{1per} }
 \end{minipage}
 \end{center}
\end{figure}

It can be seen that at large Trotter numbers the HOA method converges
with $P^{-4}$ to the quantum limit, while all other methods only converge
with $P^{-2}$. The prefactor for the EPr, however, is much smaller than for
r-EPr and PA. The value of $P$ at which convergence starts is similar in all
approaches, e.g., $k_B T P$ is larger than but in the order of
$\hbar \sqrt{k/m}$. The accuracy where HOA becomes better than EPr is 1.7\%.
For this particular model system, this value of 1.7\%
was found to be independent of
temperature. We expect it to be similar for all systems that are dominated
by harmonic interactions. We will now address the question how we have to 
increase $P$ for the various approaches if we lower $T$ and require the
relative accuracy to be constant, e.g., 1\%. The results are shown in 
Fig.~\ref{1per}. 

In order for the relative error to be constant, all methods require
that $P$ increases linearly with inverse temperature $\beta$.
The r-EPr approach, which is a little more difficult to implement than
the PA method, requires  slightly reduced Trotter numbers with respect to
$PA$. We want to emphasize that the behavior shown in Fig.~\ref{1per} 
is qualitatively similar
if the accuracy criterion for $P$ is changed, however, the stricter the 
criterion the larger the gap between  EPr and HOA. This trend can be seen
in Fig.~\ref{1per}b), where we require one time 0.1~\% accuracy instead
of 1~\% as shown in Fig.~\ref{1per}a). Only if one
is confined to the use of very small $P$,  EPr might be the better choice.
One may conclude that the optimal method depends on the desired 
accuracy. 

\subsection{Crystalline Argon}

\begin{figure}[hbtp]
\begin{center} \leavevmode
\hbox{ \epsfxsize=100mm \epsfbox{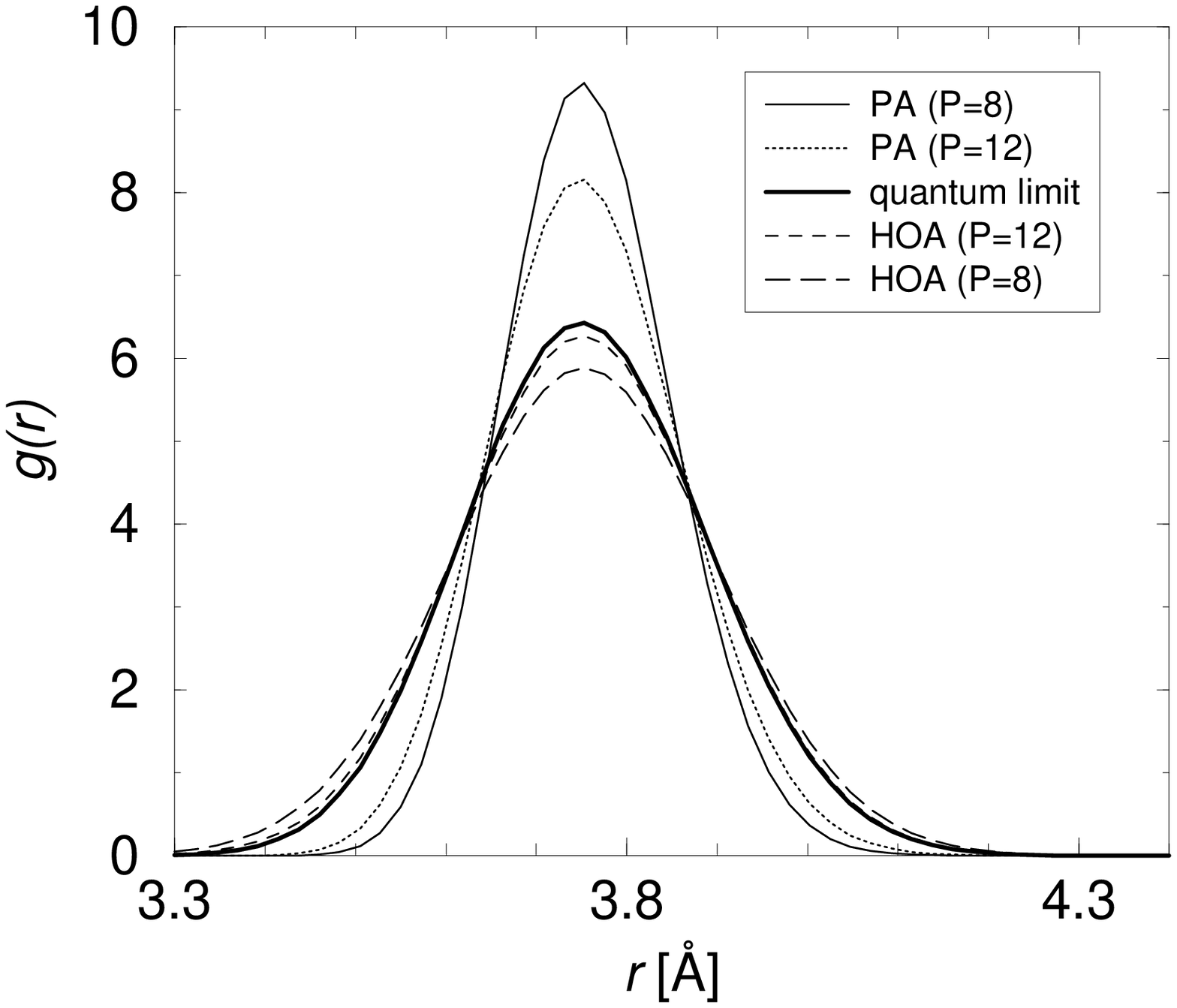} }
 \begin{minipage}{12cm}
  \caption{Pair correlation function $g(r)$ of crystalline argon at
   $T=2{\rm K}$ calculated with the PA and HOA algorithm  for
   Trotter numbers $P=8$ and $P = 12$. As a reference a 
   quasi-exact correlation function (HOA, $P=256$) is included. 
  \label{g_r} }
 \end{minipage}
 \end{center}
\end{figure}

It has been pointed out in Section~\ref{sec:HOA} that in order to calculate
radial correlation functions, it is necessary to correct the estimator 
for $g(r)$. An important question to address is how well the HOA approach 
allows one to calculate $g(r)$. In order to examine this issue we apply the 
HOA method (making use of the proper estimator, see Eq.~(\ref{eq:shift}))
to crystalline argon. The results are presented in Fig.~\ref{g_r}.

It is interesting to note that $g(r)$ is too broadened for the HOA approach
while it is too narrow using the PA algorithm. Obviously, the agreement
of the $P = 12$ HOA calculation is already very close to the quantum limit.
This is a little surprising as the product of $k_B T P$ is still far
below the thermal energy of the Debye temperature, which is about 
$T_D \approx 70$~K. 
We want to note that if the $g(r)$ are obtained without corrections,
the agreement is distinctly reduced. This might be the reason why
the convergence of $g(r)$ 
reported by Li and Broughton 
for the attractive Coulomb potential
was so slow~\cite{li87}.  In our case, omitting the corrections to the
$g(r)$ estimator leads to peaks in $g(r)$ that are even sharper than those
obtained with PA using identical Trotter numbers.

\section{Summary}

In this study we have compared the convergence to the quantum limit
for different path integral approaches.  As the convergence rate does not
depend on the specific model system (as long as the potentials are 
well-behaved), we have focused our attention to a linear
chain of harmonically coupled atoms.  We disregarded methods making explicit
use of a decomposition into a harmonic and an anharmonic part of the 
Hamiltonian among other reasons due to the unfavorable scaling of the
numerical effort with system size.

The three main approaches investigated were the so-called primitive algorithm
(PA), a method based on a higher-order approximant (HOA), and an approach in 
which  an effective potential is constructed such that the one and 
two-particle high-temperature density matrices are reproduced exactly in the 
limit of small densities. We called this approach the effective propagator 
(EPr) method. The latter approach can be further reduced
by only taking into account corrections that are local in imaginary time
leading to the r-EPr method.

We emphasized the different spirit of EPr and HOA methods: In the EPr method,
observables orthogonal in real space can be evaluated directly, whereas
the HOA method requires effective estimators. In this sense, HOA can be
seen as a non-importance sampling technique that requires the use of
effective estimators even for radial distribution functions.  The need to
alter estimators for radial distribution functions obtained in simulations
taking into account quantum effects in an effective potential had not yet been
discussed hitherto. We showed that the proper estimators lead to a very
fast convergence to the quantum limit in a simple argon crystal.

For the linear monoatomic chain investigated in this study, 
we found that the corrections based on HOA vanish with $P^{-4}$
while all other methods, PA, EPr, and r-EPr, have corrections in the
order of $P^{-2}$. The prefactor is similar for PA and r-EPr and distinctly
smaller for EPr. Both EPr and r-EPr, however, give rather accurate
results at small $P$, e.g., the relative error in the
thermal expectation value of the potential energy is smaller than
10\% in those approaches, while PA and HOA differ by nearly 100\%
at low temperatures and $P = 1$.
The Trotter number where convergence starts is similar in all approaches.
If one requires high accuracy, e.g. 1\%, one needs to increase the 
Trotter number $P$ linearly in all approaches.

What can we conclude for path integral simulations?
The use of HOA is certainly a little more (about twice) CPU time expensive 
than that of EPr, r-EPr, and PA (which all require approximataly the same
amount of computing). This is because in an HOA based simulation,
we need derivatives of the interaction potential that are one order higher than
those in simulations based on EPr, r-EPr, and PA.
However, one is rewarded with the best convergence to the quantum limit
in a HOA based path integral simulation.
While EPr also results in a
significant improvement with respect to PA, it is plagued with
a tedious, non-trivial implementation procedure. Thus HOA should be the
method of choice unless 
many-body correlations do not play a significant role like in gaseous helium.

\acknowledgments
We thank Kurt Binder for useful discussions.
Support from the BMBF through Grant 03N6015 and
from the Materialwissenschaftliche Forschungszentrum  
Rheinland-Pfalz is gratefully acknowledged.

\newpage

\begin{center} \begin{minipage}{12cm} \begin{table} 
\begin{tabular}{l|cccc}
              & PA & HOA                          & EPr               &r-EPr \\ \hline
$\tilde{k}_1$ & 0  & $(\hbar \beta k)^2 /(3mP^2)$ &$k(A-C-1)$ &  $k (A-1)$ \\
$\tilde{k}_2$ & 0  & $(\hbar \beta k)^2 /(12mP^2)$& 0                 &$0$ \\
$\tilde{\kappa}_1$ & 0 &  0                       & $-kC$             &$0$ \\
$\tilde{\kappa}_2$ & 0 &  0                       &${1 \over 2}kC$    &$0$ \\
\end{tabular} 
\caption{Expressions for the effective coupling coefficients that are 
represented in Fig.~\protect\ref{fig:illu}} \label{tab:results}
\end{table} \end{minipage} \end{center}


\newpage

\newpage

\newpage

\begin{thebibliography}{99}

\bibitem{barker79}
J.  A. Barker, 
J. Chem. Phys. {\bf 70}, 2914 (1979).

\bibitem{tuckerman93}
M. E. Tuckerman, B. J. Berne, G. J. Martyna, and M. L. Klein,
J. Chem. Phys. {\bf 99}, 2796 (1993).

\bibitem{ceperley95}
D. M. Ceperley, 
Rev. Mod. Phys. {\bf 67}, 279 (1995).

\bibitem{barrat89}
J.-L. Barrat, P. Loubeyre, and M. L. Klein,
J. Chem. Phys. {\bf 90}, 5644 (1989).

\bibitem{muser95} 
M. H. M\"user, P. Nielaba, and K. Binder,
Phys. Rev. B {\bf 51}, 2723 (1995).

\bibitem{jacobs87}
L. Jacobs, J. V. Jos{\'e}, M. A. Novotny, and A. M. Goldman,
Europhys. Lett. {\bf 3}, 1295 (1987).

\bibitem{jacobs88}
L. Jacobs, J. V. Jos{\'e}, M. A. Novotny, and A. M. Goldman,
Phys. Rev. B {\bf 38}, 4562 (1988).

\bibitem{runge92}
K. J. Runge, M. P. Surh, C. Mailhiot, and E. L. Pollock,
Phys. Rev. Lett. {\bf 69}, 3527 (1992).

\bibitem{muser96} 
M. H. M\"user and B. J. Berne,
Phys. Rev. Lett. {\bf 77}, 2638 (1996).


\bibitem{noya97}
J. C. Noya, C. P. Herrero, and R. Ramirez,
Phys. Rev. B {\bf 56} 237  (1997).

\bibitem{martonak98}
R. Martonak, W. Paul, and K. Binder,
Phys. Rev. E {\bf 57} 2425 (1998).

\bibitem{herrero01}
C. P. Herrero and R. Ramirez,
Phys Rev B {\bf 63}, 024103 (2001).

\bibitem{muser01}
M. H. M\"user,
J.  Chem. Phys. {\bf 114}, 6364 (2001).

\bibitem{candido01}
L. Candido, P. Phillips, D. M. Ceperley,
Phys. Rev. Lett. {\bf 86}, 492 (2001).

\bibitem{takahashi84}
M. Takahashi and M. Imada,
J. Phys. Soc. Jpn. {\bf 53}, 3765 (1984).

\bibitem{pollock88}
E. L. Pollock,
Comput. Phys. Commun. {\bf 52}, 49 (1988).

\bibitem{giachetti86}
R. Giachetti and V. Tognetti,
Phys. Rev. B {\bf 33}, 7647 (1986).

\bibitem{cuccoli93}
A. Cuccoli, A. Macchi, V. Tognetti, and R. Vaia,
Phys. Rev. B {\bf 47}, 14923 (1993).

\bibitem{li87}
X.-P. Li and J. Q. Broughton,
J. Chem. Phys. {\bf 86}, 5094 (1987).

\bibitem{muser97}
M. H. M{\"u}ser and B. J. Berne,
J. Chem. Phys. {\bf 107}, 571 (1997).

\bibitem{suzuki87}
M. Suzuki, in {\it Quantum Monte Carlo methods},
Springer Series in Solid-State Sciences 74 (Springer, Berlin, 1987).

\bibitem{feynman65}
R. P. Feynman and A. R. Hibbs,
{\it Quantum Mechanics and Path Integrals},
(Mc.Graw-Hill, New York, 1965).

\bibitem{feynman72}
R. P. Feynman,
{\it Statistical Mechanics},
(W. A. Benjamin, Inc. , 1972).

\bibitem{raedt83}
H. De Raedt and B. De Raedt, Phys. Rev. A {\bf 28}, 3575 (1983).

\bibitem{thirumalai84}
D. Thirumalai, R. W. Hall, and B. J. Berne,
J. Chem. Phys. {\bf 81}, 2523 (1984).

\bibitem{pollock84} 
E. L. Pollock and D. M. Ceperley, 
Phys. Rev. B {\bf 30}, 2555 (1984).

\end{thebibliography}
\end{document}